\documentclass[aps, prl,twocolumn,reprint,amsmath,amssymb,superscriptaddress,floatfix,footinbib,longbibliography]{revtex4-2}

\usepackage{epsfig}
\usepackage{graphicx}
\usepackage{epstopdf}
\usepackage[T1]{fontenc}
\usepackage[applemac]{inputenc}
\usepackage{lmodern}
\usepackage[english]{babel}
\usepackage{lipsum}
\usepackage{ae}
\usepackage{units}

\usepackage{amsmath,amssymb,bm}
\usepackage{psfrag}
\usepackage{subfigure}
\usepackage{amsthm}
\usepackage{bbm}
\usepackage{booktabs}
\usepackage{tabularx}

\usepackage{tikz}
\usetikzlibrary{arrows}

\usepackage{color}
\usepackage{url}
\usepackage[colorlinks]{hyperref}
\hypersetup{%
        plainpages=true,
        breaklinks=true,
        hypertexnames=false,
        pageanchor=true,
        colorlinks=true,
        linkcolor={blue},
        citecolor={magenta},
        urlcolor={blue},
        anchorcolor={black}
      }
\usepackage[all]{hypcap} %

\usepackage{mleftright} %

\newcommand{\figref}[1]{\mbox{Fig.~\ref{#1}}}
\newcommand{\tabref}[1]{\mbox{Table~\ref{#1}}}

\renewcommand{\eqref}[1]{\mbox{Eq.~(\ref{#1})}}
\newcommand{\figpanel}[2]{Fig.~\hyperref[#1]{\ref*{#1}(#2)}}
\newcommand{\figpanels}[3]{Fig.~\hyperref[#1]{\ref*{#1}(#2)-(#3)}}
\newcommand{\figpanelNoPrefix}[2]{\hyperref[#1]{\ref*{#1}(#2)}}

\usepackage{dcolumn}%
\usepackage{bm}%
\usepackage{nameref}
\usepackage{hyperref}
\usepackage{cleveref}
\usepackage{xcolor}
\usepackage{physics}

\newcommand{\superop}[2]{#1\{#2\}}

\begin{document}

\title{Effective Hamiltonian for an off-resonantly driven qubit-cavity system}

\author{Martin Jirlow}
\email{martin.jirlow@chalmers.se}
\affiliation{Department of Microtechnology and Nanoscience, Chalmers University of Technology, 412 96 Gothenburg, Sweden}

\author{Kunal Helambe}
\affiliation{Department of Microtechnology and Nanoscience, Chalmers University of Technology, 412 96 Gothenburg, Sweden}

\author{Axel M. Eriksson}
\affiliation{Department of Microtechnology and Nanoscience, Chalmers University of Technology, 412 96 Gothenburg, Sweden}

\author{Simone Gasparinetti}
\affiliation{Department of Microtechnology and Nanoscience, Chalmers University of Technology, 412 96 Gothenburg, Sweden}

\author{Tahereh Abad}
\email{tahereh.abad@chalmers.se}
\affiliation{Department of Microtechnology and Nanoscience, Chalmers University of Technology, 412 96 Gothenburg, Sweden}

\begin{abstract}
Accurate modeling of driven light-matter interactions is essential for quantum technologies, where natural and synthetic atoms are used to store and process quantum information, mediate interactions between bosonic modes, and enable nonlinear operations.
In systems subject to multi-tone drives, however, the theoretical description becomes challenging and existing models cannot quantitatively reproduce the experimental data.
Here, we derive an effective Hamiltonian that retains slowly rotating terms, providing a general framework for accurately describing driven dynamics across platforms.
As a concrete application, we validate the theory in circuit QED, where it quantitatively reproduces experimentally measured ac Stark shifts and captures key interactions such as two-mode squeezing and beam-splitting. %
Our results establish a broadly applicable tool to engineer driven interactions in quantum information processing platforms.
\end{abstract}

\maketitle

\paragraph*{Introduction.}

Accurate control of driven light-matter interactions is central to quantum technologies, where natural and synthetic atoms coupled to bosonic modes enable information storage, processing, mediated interactions, and nonlinear operations in platforms from circuit quantum electrodynamics (QED)~\cite{joshi2021, milul2023, krasnok2023} to trapped ions~\cite{fluehmann2019, fluehmann2020, deneeve2020} and optomechanics~\cite{shandilya2021, haug2024}.
A standard route to controlling bosonic modes %
is to couple them dispersively to an ancillary nonlinear element such as a qubit. The resulting photon-number-dependent frequency shifts enable selective addressing of Fock states~\cite{hofheinz2009, kudra2022snap, ni2023}, forming the basis of universal bosonic control.
In practice, realizing nonlinear interactions often requires multi-tone off-resonant driving of the coupled system~\cite{heeres2015,  gertler2021, ni2023, kudra2022, campagne-ibarcq2018}. While experimentally effective, such driving is theoretically challenging: conventional displaced-frame treatments~\cite{mirrahimi2014, campagne-ibarcq2018} combined with rotating-wave approximations (RWA) break down in the near-resonant regime~\cite{kudra2022}. A particularly striking failure is their inability to reproduce the measured Stark shift of the qubit.

Here, we present an effective Hamiltonian framework that extends the displaced-frame approach by retaining near-resonant contributions.
The method applies broadly; similar displacement-based approaches have also been explored in optomechanical platforms~\cite{saiko2021, rabl2011, nunnenkamp2011}, trapped ions~\cite{fluehmann2018}, and atomic ensembles~\cite{motes2017}. As a concrete application, we focus here on circuit QED in the strong dispersive regime~\cite{ofek2016, hu2019, campagne-ibarcq2020, grimm2020, Cai2021, sivak2023, lachance-quirion2024}.
We validate the model against experimental data from%
~\cite{kudra2022}, where it quantitatively reproduces ac Stark shifts and accurately captures driven interactions such as two-mode squeezing and beam-splitting.
Our results clarify the limits of conventional approximations and establish a general tool for engineering bosonic interactions in quantum information processing.

\paragraph*{System Hamiltonian.}
We consider a driven system comprising a transmon-type qubit dispersively coupled to a cavity, described by the Hamiltonian~\cite{SuppMat} %
\begin{equation}\label{eq:system_hamiltonian}
    H = H_\text{free} + H_\text{drive},
\end{equation}
where
\begin{equation} \label{eq:free_hamiltonian_fourth_order}
    \begin{gathered}
    H_\text{free} = \omega_c a^\dagger a + \omega_q b^\dagger b - \frac{E_J}{24} \left[\varphi_q (b + b^\dagger) + \varphi_c(a + a^\dagger)\right]^4,
    \end{gathered}
\end{equation}
describes a bosonic cavity mode $a$ coupled to a transmon qubit mode $b$. Here $\omega_c$ and $\omega_q$ are the renormalized mode frequencies, $E_J$ is the Josephson energy, and $\varphi_c$, $\varphi_q$ are the dimensionless zero-point flux fluctuations across the junction.

The system is driven by $N$ tones on the cavity and $M$ tones on the qubit, with drive Hamiltonian
\begin{align}
    H_\text{drive} &= \sum_{n=1}^N \varepsilon^{(n)}_c \cos\left(\omega_{cd}^{(n)} t + \theta_c^{(n)}\right) a^\dagger \notag \\
    &+ \sum_{n=1}^M \varepsilon^{(n)}_q \cos\left(\omega_{qd}^{(n)} t + \theta_q^{(n)}\right) b^\dagger + \mathrm{h.c.}, \label{eq:drive_hamiltonian}
\end{align}
where $\varepsilon_i^{(n)}$ and $\theta_i^{(n)}$ are the amplitude and phase of the $n$th drive on mode $i = q, c$.
Avoiding the conventional approximation $a^\dagger e^{-i\omega t}+\text{h.c.}$, we retain the full cosine form, which contains both positive- and negative-frequency components. In a rotating frame at the drive frequency, one component becomes stationary while the other is usually discarded as rapidly oscillating. As we show below, however, this counter-rotating contribution can combine with off-diagonal terms of the system Hamiltonian and generate effective resonant processes, making it essential to keep both.

Each drive frequency can be expressed in terms of its detuning from the corresponding mode frequency,
\begin{equation}
    \omega_{id}^{(n)} = \omega_i + \Delta_i^{(n)}, \quad \text{for } i = q, c, \label{eq:drive_frequencies}
\end{equation}
with detuning $\Delta_i^{(n)}$ from the mode frequency $\omega_i$.

Moving to a rotating frame at the cavity and qubit frequencies, and applying the RWA, the diagonal part of \eqref{eq:free_hamiltonian_fourth_order} becomes%
\begin{equation} \label{eq:rf_diagonal}
    H_\text{free}^\text{diag} = - \frac{\alpha}{2}b^{\dagger 2}b^2 - \frac{K_c}{2}a^{\dagger 2}a^2 - \chi b^\dagger b a^\dagger a,
\end{equation}
where $\alpha$ is the qubit anharmonicity, $K_c$ the cavity Kerr, and $\chi$ the dispersive coupling.
In the absence of drives, rapidly rotating terms beyond \eqref{eq:rf_diagonal} average out and can be neglected. %
In the presence of driving, however, these terms can combine with counter-rotating components of the drives \eqref{eq:drive_hamiltonian}, cancel their oscillations, and become effectively resonant in the displaced frame. As a result, contributions such as $b^{\dagger 3} b$ that rotate at $2\omega_{q}$ must be retained to capture the correct dynamics. In the next section, we show how these contributions enter the effective Hamiltonian.

\paragraph*{Main result.}

In the conventional displaced-frame treatment, one obtains an effective Hamiltonian by applying a displacement transformation that removes the linear drive terms, followed immediately by the  RWA. For clarity, we will refer to this scheme as the "Early RWA" approach~\cite{campagne-ibarcq2018, kudra2022}, since the RWA is applied at an early stage of the derivation. This procedure exposes higher-order interactions from the transmon nonlinearity, such as $a^\dagger b^\dagger$ enabling two-excitation processes or $a b^\dagger$ corresponding to excitation exchange.

In our approach, we first move into a frame rotating at the cavity and qubit frequencies, and then apply the displacement transformation. This allows certain counter-rotating contributions of the drive to survive in the displaced frame, which play a key role in the driven dynamics~\cite{SuppMat}.
We then apply the RWA only at the final step, an approach we call "Late RWA", which ensures that slowly rotating contributions at small detunings, such as the $b^{\dagger 2}b$ term in \eqref{eq:hamiltonian_nd}, are retained.
The displacement transformation is implemented by the unitary
\begin{equation}\label{eq:displacement}
    U(t) = D_q[\xi_q(t)] D_c[\xi_c(t)],
\end{equation}
where $D_q$ and $D_c$ are displacement operators for the qubit and cavity modes. The displacement amplitudes $\xi_q(t)$ and $\xi_c(t)$ are chosen to cancel the linear drive terms of \eqref{eq:drive_hamiltonian} and decompose into contributions from both co- and counter-rotating drive components,
\begin{equation}
    \xi_i(t) = \sum_{n=1}^{N_i} \left[\xi_{i,1}^{(n)}(t) + \xi_{i,2}^{(n)}(t) e^{i2\omega_i t} \right], \quad \text{for } i = q, c, \label{eq:xi_general}
\end{equation}
where the second term originates from the counter-rotating part of the cosine drives.
The explicit forms of $\xi_{i,1}^{(n)}$ and $\xi_{i,2}^{(n)}$ are given by~\cite{SuppMat}
\begin{equation}
\begin{aligned}
    \xi_{i,1}^{(n)}(t) &= \frac{\varepsilon_i^{(n)} e^{-i\theta_i^{(n)}}}{-2\Delta_i^{(n)} - i\kappa_i} e^{-i\Delta_i^{(n)} t}, \\
    \xi_{i,2}^{(n)}(t) &= \frac{\varepsilon_i^{(n)} e^{i\theta_i^{(n)}}}{4\omega_i + 2\Delta_i^{(n)} - i\kappa_i} e^{i\Delta_i^{(n)} t}, 
\end{aligned} \quad \text{for } i = q, c, \label{eq:xi_general_parts}
\end{equation}
where $\kappa_q$ and $\kappa_c$ are the qubit and cavity decay rates.
Including these rates ensures consistency with the Lindblad master equation, since the inhomogeneous transformation that defines the displacement must leave the dissipators invariant~\cite{SuppMat}. In the limit $\Delta_i^{(n)} \gg \kappa_i$, their contribution becomes negligible.
For compactness, we write $\xi_{i,1}$ and $\xi_{i,2}$ as shorthand for the sums 
$\xi_{i,1}(t) = \sum_{n=1}^{N_i} \xi_{i,1}^{(n)}(t)$ and 
$\xi_{i,2}(t) = \sum_{n=1}^{N_i} \xi_{i,2}^{(n)}(t)$, respectively, with $i=q,c$.
For brevity, we omit the explicit time dependence of these terms in what follows.

Under the assumption that the drive detunings are small enough to avoid superharmonic processes such as $b^{\dagger 3} + \mathrm{h.c.}$,
we apply \eqref{eq:displacement} to the total Hamiltonian \eqref{eq:system_hamiltonian} in the rotating frame and, after performing the RWA, obtain the effective Hamiltonian~\cite{SuppMat}
\begin{equation}
    H_\text{eff} = H_\text{diag} + H_1 + H_2.
    \label{eq:effective_hamiltonian}
\end{equation}
Here, $H_\text{diag}$ presents the diagonal part
\begin{equation}
    H_\text{diag} = \delta_q b^\dagger b + \delta_c a^\dagger a - \frac{\alpha}{2}b^{\dagger 2}b^2 - \frac{K_c}{2}a^{\dagger 2}a^2 - \chi b^\dagger b a^\dagger a,
\end{equation}
with drive-induced shifts $\delta_q = - 2\alpha|\xi_{q,1}|^2 - \chi|\xi_{c,1}|^2, $ and $\delta_c = - 2K_c|\xi_{c,1}|^2 - \chi |\xi_{q,1}|^2$.
These terms correspond to the ac Stark shifts obtained in the Early RWA treatment, but that approximation fails to reproduce the experimentally observed shifts.

The interaction part $H_1$ arises from the displacement of anharmonic and dispersive terms, and takes the form%
\begin{eqnarray}
H_1 &=& -\frac{\alpha}{2}\left(\xi_{q,1}^2b^{\dagger 2} - 2\xi_{q,1} b^{\dagger 2} b\right) - \frac{K_c}{2}\left(\xi_{c,1}^2 a^{\dagger 2} - 2\xi_{c,1} a^{\dagger 2}a\right) \notag\\
&-& \chi\left(\xi_{q,1} \xi_{c,1} b^\dagger a^\dagger + \xi_{q,1}^* \xi_{c,1} b a^\dagger - \xi_{c,1} b^\dagger b a^\dagger - \xi_{q,1}b^\dagger a^\dagger a\right)\notag \\
&+& \left(\alpha \xi_{q,1} |\xi_{q,1}|^2 + \chi \xi_{q,1} |\xi_{c,1}|^2\right)b^\dagger \notag \\
&+& \left(K_c \xi_{c,1} |\xi_{c,1}|^2 + \chi \xi_{c,1} |\xi_{q,1}|^2\right)a^\dagger + \text{h.c.}. \label{eq:hamiltonian_nd}
\end{eqnarray}
Here we see, for example, that making the two-photon term $b^\dagger a^\dagger$ resonant requires canceling the time dependence in $\xi_{q,1}\xi_{c,1}$, which is achieved by choosing opposite detunings, $\Delta_c^{(n)} = -\Delta_q^{(n)} = \Delta$ in \eqref{eq:xi_general_parts}. %
The remaining terms then rotate at frequencies that are integer multiples of $\Delta$. When $1/\Delta$ is similar to or larger than the dynamical timescale, these terms remain slowly varying and contribute significantly to the system evolution. In \cite{you2024}, only the cavities are displaced, which does not capture the effects of off-resonance qubit driving.

Finally, $H_2$ contains contributions from counter-rotating drive terms, including terms oscillating at $2\omega_q$, $2\omega_c$ in the rotating frame, or higher-order mixing frequencies. These become relevant through the time dependence of $\xi_{i,2}$:
\begin{align}
H_2 &= \alpha\xi_{q,2}^* b^{\dag 2} b - K_c \xi_{c,2}^* a^{\dag 2}a - \frac{\chi}{6}\left(\xi_{q,2}^* \xi_{c,2}^* b^\dag a^\dag + \xi_{q,2}\xi_{c,2}^* b a^\dag\right) \notag\\
&+ \frac{\chi}{6}\left(\xi_{c,2}^*b^\dag b a^\dag + \xi_{q,2}^*b^\dag a^\dag a \right) + \text{h.c.}. \label{eq:hamiltonian_cr}
\end{align}
The counter-rotating contribution \eqref{eq:hamiltonian_cr} provides a small but non-negligible correction to \eqref{eq:hamiltonian_nd}. It originates from rapidly oscillating terms in the Josephson cosine expansion whose rotation is canceled by the counter-rotating components of the drive fields.

Equations (\ref{eq:effective_hamiltonian})-(\ref{eq:hamiltonian_cr}) constitute \textit{the main result} of this letter: an effective Hamiltonian describing the dynamics of a cavity-qubit system under off-resonant multi-tone driving, including frequency shifts, induced interactions, and counter-rotating corrections.

\paragraph*{AC Stark shift as a probe of model accuracy.}

\begin{figure}
    \centering
    \includegraphics{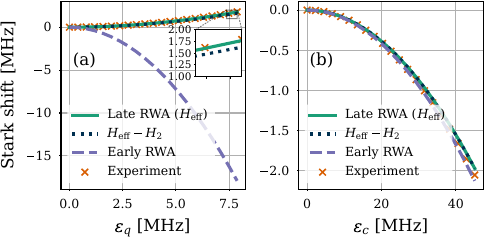}
    \caption{Ac Stark shift of the qubit frequency under (a) a qubit drive at $\omega_q + \Delta_q$ ($\Delta_q = -20~\text{MHz}$) and (b) a cavity drive at $\omega_c + \Delta_c$ ($\Delta_c = 18.5~\text{MHz}$), as functions of drive amplitudes $\varepsilon_q$ and $\varepsilon_c$, respectively. 
    Simulations using $H_\text{eff}$ obtained from the Late RWA method, show excellent agreement with experimental data, while Early RWA fails to reproduce the correct sign and magnitude in (a). The inset in (a) highlights the noticeable contribution of the correction $H_2$.}
    \label{fig:stark-shifts}
\end{figure}

We validate the effective Hamiltonian \eqref{eq:effective_hamiltonian} by comparing its predictions for the qubit ac Stark shift in the coupled qubit-cavity system with experimental data. Single-tone drives are applied either to the qubit or the cavity: the qubit is driven red-detuned at $\omega_q + \Delta_q$ ($\Delta_q < 0$), and the cavity blue-detuned at $\omega_c + \Delta_c$ ($\Delta_{c} > 0$), with the specific detunings given in the caption of \figref{fig:stark-shifts}. Simulations use the parameters of~\cite{kudra2022}.

The Stark shift is defined as the change in the dressed qubit frequency extracted from the spectrum of $H_\text{eff}$ \eqref{eq:effective_hamiltonian} relative to its undriven value. 
For single-tone drives, $H_\text{eff}$ can be made time-independent by moving into an additional rotating frame at the detuning $\Delta_i$. Since we are already in the frame rotating at $\omega_i$, this extra transformation effectively brings us into the interaction picture at the drive frequency $\omega_i+\Delta_i$. In this frame, the phase rotation of $\xi_{i,j}$ ($i = q,c$, $j = 1,2$) is canceled, yielding well-defined eigenstates and eigenenergies.
Figure \figpanelNoPrefix{fig:stark-shifts}{a} shows the shift under qubit driving, with excellent agreement between model and experiment to within $\sim 20$ kHz (1.3\% error). Figure \figpanelNoPrefix{fig:stark-shifts}{b} shows the cavity-drive case, with deviations below $\sim 81$ kHz (3.9\% error).
Importantly, the effective Hamiltonian obtained from the Late RWA approach reproduces both the \emph{magnitude and sign} of the shift, unlike the Early RWA which predicts a negative shift. The correction $H_2$, although small, makes a visible difference in \figpanel{fig:stark-shifts}{a}.

\begin{figure}
    \centering
    \includegraphics[width=3.38in]{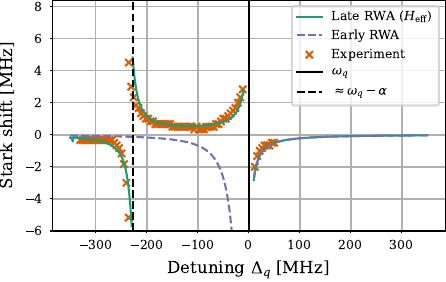}
    \caption{Qubit Stark shift versus detuning for a fixed qubit drive amplitude $\varepsilon_q = 7.63$ MHz. The Late RWA approach ($H_\text{eff}$) captures both the sign change across resonance and the transition near $\Delta_q \approx -\alpha$, features that are missed by the Early RWA method. The system parameters are listed in Table~S1~\protect\cite{SuppMat}.}
    \label{fig:shift_vs_detuning}
\end{figure}

To understand this discrepancy for $\Delta_q < 0$, note that earlier models neglect terms rotating at frequencies $\sim \Delta_q$, which is only valid when $|\Delta_q|$ is large. At small detuning, these terms rotate slowly and crucially affect the Stark shift. Figure~\ref{fig:shift_vs_detuning} shows the Stark shift versus detuning, revealing a sign change across resonance and a critical transition around $\Delta_q \approx -\alpha$, due to an avoided level crossing between the qubit $\ket{e}$ and $\ket{f}$ states in the rotating frame. This feature, consistent with experiment, is beyond the reach of previous approximations~\cite{kudra2022, campagne-ibarcq2018}.

Finally, the model also captures the Stark shift of the cavity induced by qubit driving (see Supplement~\cite{SuppMat}), which is much smaller ($\sim 50$ kHz) due to the cavity's near-harmonic spectrum. %

We now demonstrate the utility of our effective Hamiltonian by analyzing two fundamental interactions that arise when driving both the qubit and cavity with two tones: two-mode squeezing and beam-splitting.

\paragraph*{Two-mode squeezing.}

This regime involves two-photon creation processes of the form $a^\dagger b^\dagger$, enabling entanglement generation between the qubit and cavity modes~\cite{campagne-ibarcq2018, kudra2022}. A notable example is the selective number-dependent arbitrary-phase photon-addition (SNAPPA) protocol~\cite{kudra2022}, which employs off-resonant drives to realize number-selective interactions. In SNAPPA, a single qubit drive at $\Delta_q = -\Delta$ is combined with a multi-tone cavity drive at $\Delta_c^{(n)} = \Delta - (n + 1)\chi$, where $n$ spans the photon numbers from a chosen parity subspace (e.g., all odd $n$), $\chi$ is the dispersive shift, and $\Delta$ is a small detuning on the order of tens of MHz. These drives coherently induce transitions $\ket{n}_c \ket{g}_q \rightarrow \ket{n+1}_c\ket{e}_q$, thereby adding a photon conditioned on photon number parity, while simultaneously exciting the qubit and imprinting a programmable phase dependent on the relative phases of the drive fields. Such number- and parity-selective operations are key resources for bosonic error correction~\cite{kudra2022, gertler2021}.

\begin{figure}
    \centering
    \includegraphics{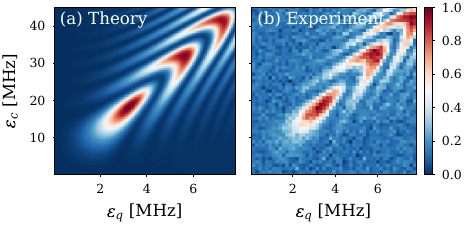}
    \caption{Qubit population under the two-mode squeezing transition $\ket{0}_c \ket{g}_q \rightarrow \ket{1}_c \ket{e}_q$, as a function of drive amplitudes $\varepsilon_q$ and $\varepsilon_c$, after a gate time $\tau = 4.2~\mu$s. (a) Simulation using the effective Hamiltonian \eqref{eq:effective_hamiltonian}. (b) Experimental data from~\protect\cite{kudra2022}.}
    \label{fig:chevrons}
\end{figure}

We simulate $H_\text{eff}$, \eqref{eq:effective_hamiltonian}, using the system parameters as in~\cite{kudra2022}, with single-tone drives on the qubit and cavity at amplitudes $\varepsilon_q$ and $\varepsilon_c$. %
Drive frequencies are chosen to implement the SNAPPA transition $\ket{0}_c\ket{g}_q \rightarrow \ket{1}_c\ket{e}_q$ (i.e., $n=0$). Figure~\ref{fig:chevrons}(a) shows the simulated qubit population, as $\varepsilon_q$ and $\varepsilon_c$ are varied. The resulting interference pattern closely matches the experimental data in Fig.~\ref{fig:chevrons}(b), confirming that Late RWA captures the relevant dynamics. Additional simulations in~\cite{SuppMat} show qubit populations for higher-order two-mode squeezing transitions, where $H_\text{eff}$ reproduces the experimental chevrons while the Early RWA model shows clear discrepancies.

\paragraph*{Beam splitting interaction.}

\begin{figure}
    \centering
    \includegraphics{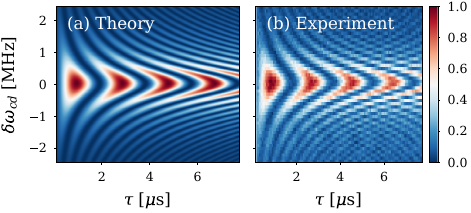}
    \caption{Qubit population under the beam splitting transition 
    $\ket{1}_c \ket{g}_q \leftrightarrow \ket{0}_c \ket{e}_q$, 
    as a function of gate time $\tau$ and cavity drive frequency offset $\delta\omega_{cd}$. 
    The drive frequencies are $\omega_{qd} = \omega_q + \Delta_q$ and 
    $\omega_{cd} = \omega_c + \Delta_c + \nu_\text{corr} + \delta\omega_{cd}$, 
    with $\Delta_q = \Delta_c = -50$~MHz and 
    $\nu_\text{corr} \approx -5.02$~MHz an experimental correction that centers the pattern. 
    (a) Simulation using the effective Hamiltonian \eqref{eq:effective_hamiltonian}. 
    (b) Experimental data, taken with the parameters of Table~S1 in~\protect\cite{SuppMat}.}
    \label{fig:beam_splitting}
\end{figure}

This regime corresponds to another two-photon process in which an excitation is swapped between the qubit and cavity modes, and is commonly used to implement controlled SWAP gates~\cite{chapman2023, campagne-ibarcq2018, gao2019}.
It can be driven using a qubit and cavity tone with matched detunings, $\Delta_q = \Delta_c = \Delta$, such that $\omega_{qd} - \omega_{cd} = \omega_q - \omega_c$. This condition selectively drives the conversion $\ket{n}_c\ket{e}_q \leftrightarrow \ket{n+1}_c\ket{g}_q$.  
Figure \figpanelNoPrefix{fig:beam_splitting}{a} shows the simulated qubit population for the beam-splitting transition, starting from the initial state $\ket{1}_c\ket{g}_q$, as a function of gate time $\tau$ and cavity drive frequency offset $\delta\omega_{cd}$.
The drives are set with $\Delta = -50$~MHz, and $\omega_{cd} = \omega_c + \Delta + \nu_\text{corr} + \delta\omega_{cd}$, where $\nu_\text{corr} \approx -5.02$~MHz is an experimental correction that centers the beam splitting pattern. The simulation uses the system parameters listed in Table~S1 of~\cite{SuppMat}. The simulated dynamics closely match the experimental data in Fig.~\figpanelNoPrefix{fig:beam_splitting}{b}, demonstrating that the effective Hamiltonian provides an accurate description of the beam-splitting interaction.

\paragraph*{Outlook.}
We derived an effective Hamiltonian for a driven cavity-transmon system that goes beyond the conventional displaced-frame treatment, which we refer to as the Early RWA method.
In that approach, the RWA is applied immediately after the displacement transformation.
In contrast, our Late RWA method first moves into a rotating frame, then performs the displacement, and only applies the RWA at the end. Deferring the RWA in this way preserves counter-rotating drive contributions in the displaced frame.
The model applies to off-resonant multi-tone driving, with particular importance in the near-resonant regime, where the Early RWA method fails.
It reproduces experimental features, such as the qubit and cavity ac Stark shifts, and accurately captures beam-splitting and two-photon processes.

Our effective Hamiltonian serves as a flexible tool for simulating systems under arbitrary multitone driving on both qubit and cavity. This enables precise predictions of cumulative Stark effects from complex drive spectra. Such capability is particularly valuable for high-fidelity control, where even small, Fock-dependent shifts, especially in the cavity, can be used to sculpt the transition landscape and mitigate unwanted nonlinearities~\cite{huang2025}.
While we focused here on the fundamental transmon transition, the same approach naturally extends to any anharmonic oscillator, including other qubit types such as the fluxonium~\cite{koch2009, manucharyan2009, nguyen2019, nie2025}, phase~\cite{martinis2009}, and flux qubits~\cite{orlando1999, mooij1999, koch2007}.

Looking forward, extending the method to include superharmonic processes could open the door to modeling higher-order transitions such as \(\ket{0f} \leftrightarrow \ket{1g}\)~\cite{Magnard2018FastReset}, and to applications ranging from dissipation engineering via lossy buffer modes~\cite{gertler2021} to on-demand microwave photon generation in parametrically driven systems~\cite{yang2025, yang2018deterministic}. We expect this framework to become a practical tool for advancing control techniques in circuit QED platforms.

\paragraph*{Acknowledgements.}
We acknowledge useful discussions with G\"{o}ran Johansson, Marina Kudra, and Alberto Del \'{A}ngel Medina.
All simulations and visualizations were performed using QuTiP~\cite{johansson2012, johansson2013}, NumPy~\cite{harris2020} and Matplotlib~\cite{hunter2007}.
We acknowledge support from the Knut and Alice Wallenberg Foundation through the Wallenberg Centre for Quantum Technology (WACQT). S.G. acknowledges financial support from the European Research Council (Grant No. 101041744 ESQuAT). External interest disclosure: S.G. is a co-founder and equity holder in Sweden Quantum AB.

T.A. conceived and planned the project.
M.J. developed the theoretical model, performed the simulations, and analyzed the results.
K.H. performed the experiments and analyzed the data with input from A.M.E..
S.G. supported the experimental aspects of the work and provided feedback on the manuscript.
M.J. and T.A. wrote the manuscript with feedback from all authors.
T.A. supervised the project.

\bibliography{references}

\onecolumngrid
\clearpage
\setcounter{section}{0}
\renewcommand{\thesection}{\alph{section}}
\setcounter{equation}{0}
\renewcommand{\theequation}{S\arabic{equation}}
\setcounter{figure}{0}
\renewcommand{\thefigure}{S\arabic{figure}}
\setcounter{table}{0}
\renewcommand{\thetable}{S\arabic{table}}

\setcounter{page}{1}

\section{Derivation of the Effective Hamiltonian}\label{app:ham_derivation}
Here we present the details of the derivation of the effective Hamiltonian \eqref{eq:effective_hamiltonian} using the displacement-transformation formalism. The system consists of a three-dimensional cavity capacitively coupled to a transmon qubit, described by the Hamiltonian
\begin{equation}
    H_\text{free} = \Tilde{\omega}_c a^\dagger a + \Tilde{\omega}_q b^\dagger b - E_J \cos\mleft(\varphi_q (b + b^\dagger)\mright) - g(a - a^\dagger)(b - b^\dagger),
\end{equation}
where $a$ ($a^\dagger$) and $b$ ($b^\dagger$) are the annihilation (creation) operators for the cavity and transmon modes, $\Tilde{\omega}_c$, $\Tilde{\omega}_q$ are the bare linear frequencies, $E_J$ is the Josephson energy, $\varphi_q$ is the dimensionless zero-point flux fluctuation across the junction for the qubit mode, 
and $g$ denotes the capacitive coupling strength between cavity and qubit.

The capacitive coupling term can be eliminated by moving into the normal-mode basis~\cite{malekakhlagh2020}, yielding
\begin{equation}\label{app:eq:initial_hamiltonian}
    H_\text{free} = \bar{\omega}_c a^\dagger a + \bar{\omega}_q b^\dagger b - E_J\cos\left(\varphi_a(a + a^\dagger) + \varphi_q(b + b^\dagger)\right),
\end{equation}
where $\bar{\omega}_c$ and $\bar{\omega}_q$ denote the normal-mode frequencies, and $\varphi_{c,q}$ are the zero-point flux fluctuations across the Josephson junction.
Assuming the system operates in the dispersive regime, $|\Tilde{\omega}_q - \Tilde{\omega}_c| \gg g$, the normal-mode frequencies are close to the bare values, $\bar{\omega}_c \approx \Tilde{\omega}_c$ and $\bar{\omega}_q \approx \Tilde{\omega}_q$. Expanding the cosine in \eqref{app:eq:initial_hamiltonian} to fourth order and neglecting higher-order terms yields
\begin{equation}\label{app:eq:free_hamiltonian}
    H_\text{free} = \omega_c a^\dagger a + \omega_q b^\dagger b - \frac{E_J}{4!}\left[\varphi_a(a + a^\dagger) + \varphi_q(b + b^\dagger)\right]^4,
\end{equation}
where the quadratic term renormalizes the cavity and qubit frequencies. This corresponds to \eqref{eq:free_hamiltonian_fourth_order} in the main text.

We now include external drives with multiple frequency components applied to both the qubit and the cavity. %
The drive Hamiltonian reads
\begin{equation}
    H_\text{drive} = \sum_{n}\varepsilon_q^{(n)} \cos\left(\omega_{qd}^{(n)} t + \theta^{(n)}_q\right)b^\dag + \sum_{n} \varepsilon_c^{(n)} \cos\left(\omega_{cd}^{(n)} t + \theta^{(n)}_c\right)a^\dag + \mathrm{H.c.},
\end{equation}
which, using complex exponentials, can be expressed as
\begin{equation}
    H_\text{drive} = \sum_{n} \frac{\varepsilon_q^{(n)}}{2}\left(e^{-i\left(\omega_{qd}^{(n)} t + \theta_q^{(n)}\right)} + e^{i\left(\omega_{qd}^{(n)} t + \theta_q^{(n)}\right)}\right)b^\dagger  + \sum_{n} \frac{\varepsilon^{(n)}_c}{2}\left(e^{-i\left(\omega^{(n)}_{cd} t + \theta_c^{(n)}\right)} + e^{i\left(\omega^{(n)}_{cd} t +\theta_c^{(n)}\right)}\right)a^\dagger + \mathrm{H.c.},
\end{equation}
So that the total Hamiltonian is 
\begin{equation}
H = H_\text{free} + H_\text{drive}.
\end{equation}

The system dynamics are described by the Lindblad master equation
\begin{equation}\label{app:eq:lindblad}
    \frac{\dd \rho}{\dd t} = - i \left[H, \rho\right] + \sum_i \kappa_i\superop{\mathcal{D}}{L_i}\rho,
\end{equation}
with dissipation superoperators
\begin{equation}\label{app:eq:dissipator}
    \superop{\mathcal{D}}{L_i}\rho = L_i \rho L_i^\dag - \frac{1}{2}\left\{L_i^\dag L_i, \rho\right\}.
\end{equation}
where $L_i$ are the jump operators and $\kappa_i$ their associated rates.  
For the qubit-cavity system we take
\begin{equation}
    L_i \in \left\{a, b, b^\dag b\right\},
\end{equation}
corresponding respectively to cavity photon loss, qubit relaxation, and qubit dephasing, with rates $\kappa_c$, $\kappa_q$, and $\kappa_d$.

To derive the effective Hamiltonian, we move into a frame rotating at the bare resonance frequencies of the cavity and qubit, $\omega_c$ and $\omega_q$.
This leaves the dissipation superoperators unchanged but transforms the Hamiltonian.
In this rotating frame, the annihilation operators become
\begin{align}
    a^\prime = a e^{-i\omega_c t}, \\
    b^\prime = b e^{-i\omega_q t},
\end{align}
leading to the transformed Hamiltonian
\begin{equation}\label{app:eq:rf_ham}
    H^\prime = H_\text{free}^\prime + H_\text{drive}^\prime.
\end{equation}

The transformed free Hamiltonian is
\begin{equation}
    \begin{aligned}
    H_\text{free}^\prime = - \frac{E_J}{4!}\left[\varphi_a \left(a^\prime + a^{\prime \dag}\right) + \varphi_q\left(b^\prime + b^{\prime \dag}\right)\right]^4,
    \end{aligned}
\end{equation}
while the drive Hamiltonian in the rotating frame takes the form
\begin{equation}
    H_\text{drive}^\prime = \sum_n\frac{\varepsilon_q^{(n)}}{2}\left(e^{-i\left(\Delta_q^{(n)}t +\theta_q^{(n)}\right)} + e^{i\left(\left(2\omega_q + \Delta_{q}^{(n)}\right)t + \theta_q^{(n)}\right)}\right)b^\dag + \sum_n\frac{\varepsilon^{(n)}_c}{2}\left(e^{-i\left(\Delta_c^{(n)}t + \theta_c^{(n)}\right)} + e^{i\left(\left(2\omega_c + \Delta_{c}^{(n)}\right)t + \theta_c^{(n)}\right)}\right)a^\dag + \mathrm{H.c.},
\end{equation}
where we use $\omega_{id}^{(n)} = \omega_i + \Delta_i^{(n)}$ for $i = q,c$.

To capture the nonlinear effects of off-resonant driving, we apply a displacement transformation to both qubit and cavity modes, introducing time-dependent displacement amplitudes $\xi_q(t)$ and $\xi_c(t)$. The displaced operators are defined as
\begin{align}
    \Tilde{a} &= a^\prime - \xi_c, \label{app:eq:disp_op_1}\\
    \Tilde{b} &= b^\prime - \xi_q,\label{app:eq:disp_op_4}%
\end{align}
together with the inhomogeneous Hamiltonian shift
\begin{equation}
    H \rightarrow H + i\frac{\kappa_q}{2}\left(\xi_q b^\dag - \xi_q^* b\right) + i\frac{\kappa_c}{2}\left(\xi_c a^\dag - \xi_c^* a\right),
\end{equation}
which leaves the Lindblad dissipators $\mathcal{D}\{a\}$ and $\mathcal{D}\{b\}$ invariant~\cite{Breuer2002-kb}.
Substituting into \eqref{app:eq:rf_ham} yields
\begin{equation}
      \Tilde{H} = \Tilde{H}_\text{free} + \Tilde{H}_\text{drive}, \label{app:eq:disp_ham}  
\end{equation}
with
\begin{align}
    \Tilde{H}_\text{free} &= - \frac{E_J}{4!}\left[\varphi_a \left(\Tilde{a} + \Tilde{a}^\dag\right) + \varphi_q\left(\Tilde{b} + \Tilde{b}^\dag\right)\right]^4,\label{app:eq:free_displaced}\\
        \Tilde{H}_\text{drive} &= \left(i\frac{\partial \xi_q}{\partial t} + i\frac{\kappa_q}{2}\xi_q + \sum_n\frac{\varepsilon^{(n)}_q}{2}\left(e^{-i\Delta_q^{(n)}t}e^{-i\theta_q^{(n)}} + e^{i\left(2\omega_q + \Delta_{q}^{(n)}\right)t}e^{i\theta_q^{(n)}}\right)\right)b^\dag + \notag \\
        &+ \left(i\frac{\partial \xi_c}{\partial t} + i\frac{\kappa_c}{2}\xi_c + \sum_n\frac{\varepsilon^{(n)}_c}{2}\left(e^{-i\Delta_c^{(n)}t}e^{-i\theta_c^{(n)}} + e^{i\left(2\omega_c + \Delta_{c}^{(n)}\right)t}e^{i\theta_c^{(n)}}\right)\right)a^\dag + \mathrm{H.c.},
\label{app:eq:drive_displaced}
\end{align}

To account for multiple drive tones, we decompose the displacements as
\begin{gather}
    \xi_q(t) = \sum_n \xi_q^{(n)}(t), \\
    \xi_c(t) = \sum_n \xi_c^{(n)}(t), 
\end{gather}
which gives the drive Hamiltonian
\begin{align}
    \Tilde{H}_\text{drive} &= \mleft(\sum_n \mleft( i\frac{\partial \xi_q^{(n)}}{\partial t} + i\frac{\kappa_{q}}{2}\xi_q^{(n)} \mright)+ \frac{\varepsilon_q^{(n)}}{2}\left(e^{-i \Delta_q^{(n)} t}e^{-i\theta_q^{(n)}} + e^{i\left(2\omega_q + \Delta_{q}^{(n)}\right)t}e^{i\theta_q^{(n)}}\right)\mright)b^\dag + \notag\\
    &+ \left(\sum_n \mleft( i\frac{\partial \xi^{(n)}_c}{\partial t} + i\frac{\kappa_c}{2}\xi^{(n)}_c \mright) + \frac{\varepsilon^{(n)}_c}{2}\left(e^{-i\Delta_c^{(n)}t}e^{-i\theta_c^{(n)}} + e^{i\left(2\omega_c + \Delta_{c}^{(n)}\right)t}e^{i\theta_c^{(n)}}\right)\right)a^\dag + \mathrm{H.c.}
\label{app:eq:drive_displaced_sum}
\end{align}

The displacement amplitudes $\xi_q^{(n)}$ and $\xi_c^{(n)}$ are chosen to cancel the drive terms in \eqref{app:eq:drive_displaced_sum}, obtained by setting the corresponding coefficients to zero. This condition enforces $\Tilde{H}_\text{drive} = 0$, and the resulting solutions take the form 
\begin{align}
    \xi_q^{(n)}(t) &= \xi_{q,1}^{(n)}(t) + \xi_{q,2}^{(n)}(t) e^{i2\omega_q t}, \\
    \xi_c^{(n)}(t) &= \xi_{c, 1}^{(n)}(t) + \xi_{c,2}^{(n)}(t) e^{i2\omega_c t},
\end{align}
The individual components are
\begin{align}
    \xi_{q,1}^{(n)}(t) &= \frac{\varepsilon_q^{(n)} e^{-i\theta_q^{(n)}}}{-2\Delta_q^{(n)} - i\kappa_{q}}e^{-i\Delta_q^{(n)} t}, \\
    \xi_{q,2}^{(n)}(t) &= \frac{\varepsilon_q^{(n)} e^{i\theta_q^{(n)}}}{4\omega_q + 2\Delta_q^{(n)} - i\kappa_{q}} e^{i\Delta_q^{(n)} t}, \\
    \xi_{c,1}^{(n)}(t) &= \frac{\varepsilon^{(n)}_c e^{-i\theta^{(n)}_c}}{-2\Delta_c^{(n)} - i\kappa_c} e^{-i \Delta_c^{(n)}t}, \\
    \xi_{c,2}^{(n)}(t) &= \frac{\varepsilon_c^{(n)} e^{i\theta^{(n)}_c}}{4\omega_c + 2\Delta_c^{(n)} - i\kappa_c}e^{i \Delta_c^{(n)}t}.
\end{align}
For brevity, we will write $\xi_{i,1}(t) = \sum_n \xi_{i,1}^{(n)}(t)$ and $\xi_{i,2}(t) = \sum_n \xi_{i, 1}^{(n)}(t)$ respectively for $i = q,c$, and omit the explicit time dependence.

To evaluate \eqref{app:eq:disp_ham}, which now reduces to \eqref{app:eq:free_displaced}, we substitute the displaced operators $\tilde{a}$ and $\tilde{b}$ from \eqref{app:eq:disp_op_1} and \eqref{app:eq:disp_op_4}, expressed in terms of the original operators $a$, $b$ and the time-dependent displacement amplitudes $\xi_c(t)$ and $\xi_q(t)$.

To apply the rotating wave approximation (RWA), we assume drive detunings are small enough to neglect superharmonic processes, such as three-photon transitions ($b^{\dag 3}$) and similar terms.

Expanding the quartic interaction in \eqref{app:eq:free_displaced} and applying the RWA to remove rapidly oscillating terms, we obtain the effective Hamiltonian
\begin{equation} \label{app:eq:ham}
    H_\text{eff} = H_\text{diag} + H_\text{int}^{(1)} + H_\text{int}^{(2)}, 
\end{equation}
which corresponds to \eqref{eq:effective_hamiltonian} in the main text.
The diagonal part is
\begin{equation}
    H_\text{diag} = \delta_q b^\dagger b + \delta_c a^\dagger a - \frac{\alpha}{2}b^{\dagger 2}b^2 - \frac{K_c}{2}a^{\dagger 2}a^2 - \chi b^\dagger b a^\dagger a,
\end{equation}
with drive-induced frequency shifts
\begin{align}
    \delta_q &= - 2\alpha|\xi_{q,1}|^2 - \chi|\xi_{c,1}|^2, \\
    \delta_c &= - 2K_c|\xi_{c,1}|^2 - \chi |\xi_{q,1}|^2.
\end{align}

The off-diagonal interaction terms are
\begin{align}
H_\text{int}^{(1)} &= -\frac{\alpha}{2}\left(\xi_{q,1}^2b^{\dagger 2} - 2\xi_{q,1} b^{\dagger 2} b\right) - \frac{K_c}{2}\left(\xi_{c,1}^2 a^{\dagger 2} - 2\xi_{c,1} a^{\dagger 2}a\right) \notag \\
&- \chi\left(\xi_{q,1} \xi_{c,1} b^\dagger a^\dagger + \xi_{q,1}^* \xi_{c,1} b a^\dagger - \xi_{c,1} b^\dagger b a^\dagger - \xi_{q,1}b^\dagger a^\dagger a\right) \notag \\
&+ \left(\alpha \xi_{q,1} |\xi_{q,1}|^2 + \chi \xi_{q,1} |\xi_{c,1}|^2\right)b^\dagger + \left(K_c \xi_{c,1} |\xi_{c,1}|^2 + \chi \xi_{c,1} |\xi_{q,1}|^2\right)a^\dagger + \text{h.c.},
\end{align}
while the counter-rotating contributions are
\begin{equation}
H_\text{int}^{(2)} = \alpha\xi_{q,2}^* b^{\dag 2} b - K_c \xi_{c,2}^* a^{\dag 2}a - \frac{\chi}{6}\mleft(\xi_{q,2}^* \xi_{c,2}^* b^\dag a^\dag + \xi_{q,2}\xi_{c,2}^* b a^\dag\mright) + \frac{\chi}{6}\mleft(\xi_{c,2}^*b^\dag b a^\dag + \xi_{q,2}^*b^\dag a^\dag a \mright) + \text{h.c.}.
\end{equation}

The dynamics of the open system are then described by the Lindblad master equation
\begin{equation}
    \frac{\dd \rho}{\dd t} = -i\comm{H_\mathrm{eff}}{\rho} + \kappa_c\superop{\mathcal{D}}{a}\rho + \kappa_q\superop{\mathcal{D}}{b}\rho + \kappa_d\superop{\mathcal{D}}{(b^\dag - \xi_q^*)(b - \xi_q)}\rho.
\end{equation}
where the last dissipator corresponds to qubit dephasing, which transforms under the displacement as $(b^\dag - \xi_q^*)(b - \xi_q)$.

\section{Hamiltonian parameters}\label{app:parameters}

The system parameters used in the simulations of the beam splitting interaction are listed in \tabref{tab:system_parameters}.
\begin{table}[h]
    \centering
    \begin{tabular*}{0.7\textwidth}{@{\extracolsep{\fill}}c c c}
        \hline\hline
        Parameter & Symbol & Value (MHz)  \\
        \hline
        Qubit frequency & $\omega_q/2\pi$ & 5311 \\
        Cavity frequency & $\omega_c/2\pi$ & 3579 \\
        Dispersive shift & $\chi/2\pi$ & 1.923 \\
        Cavity Kerr & $K_c/2\pi$ & 0.0022 \\
        Qubit anharmonicity & $\alpha/2\pi$ & 229.9 \\
        \hline\hline
    \end{tabular*}
    \caption{System parameters used in the simulations of the beam splitting interaction.} 
    \label{tab:system_parameters}
\end{table}

\section{Higher order transitions in two-mode squeezing}

\begin{figure}
    \centering
    \includegraphics{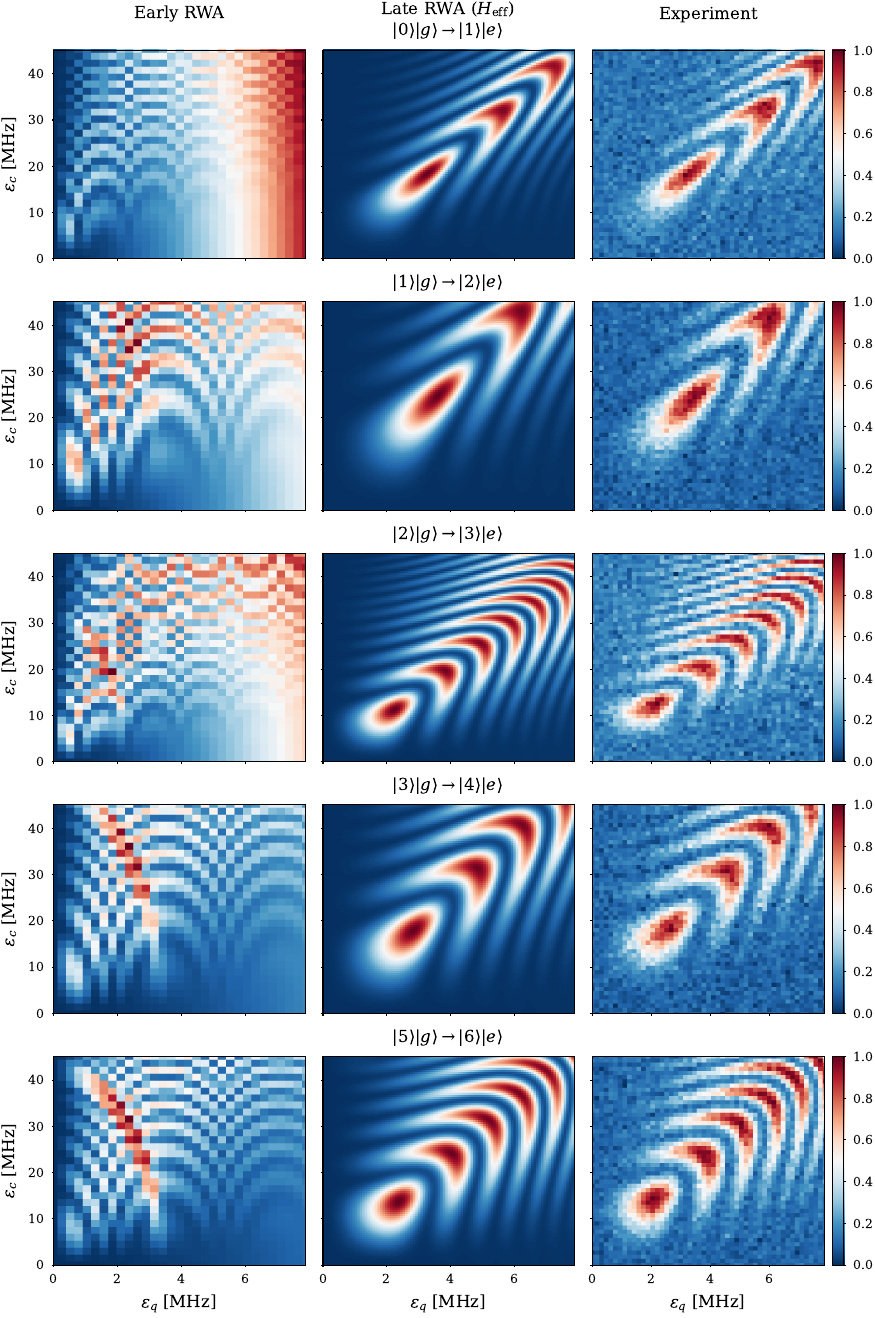}
    \caption{Qubit populations under the two-mode squeezing transition $\ket{n}_c \ket{g}_q \rightarrow \ket{n+1}_c\ket{e}_q$ as a function of drive amplitudes $\varepsilon_q$ and $\varepsilon_c$, for $n \in \{0, 1, 2, 3, 5\}$. From left to right we see (left) numerical simulation using Early RWA, (center) numerical simulation using Late RWA (\eqref{eq:effective_hamiltonian}), and (right) experimental data from~\protect\cite{kudra2022}.}
    \label{fig:chevron_combined}
\end{figure}

To further highlight the discrepancy between the conventional displaced-frame approach and experimental results at small detunings, we present here the qubit population under two-mode squeezing, analogous to \figref{fig:chevrons} of the main text. Figure \ref{fig:chevron_combined} shows the results for different photon-number states $n$ in the transition $\ket{n}_c \ket{g}_q \rightarrow \ket{n + 1}_c\ket{e}_q$, using detunings $\Delta_c = -\Delta_q = 20$ and $30$~MHz.

\section{Cavity ac Stark shifts}

\begin{figure}
    \centering
    \includegraphics[width=0.5\linewidth]{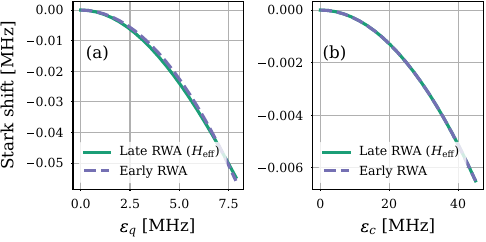}
    \caption{Ac Stark shift of the cavity frequency under (a) a qubit drive at $\omega_q + \Delta_q$ ($\Delta_q = -20~\text{MHz}$) and (b) a cavity drive at $\omega_c + \Delta_c$ ($\Delta_c = 18.56~\text{MHz}$), as functions of drive amplitudes $\varepsilon_q$ and $\varepsilon_c$, respectively.}
    \label{fig:cavity_stark_shifts}
\end{figure}

For completeness, we also show the Stark shift of the cavity predicted by our model in Fig.~\ref{fig:cavity_stark_shifts}. Because the cavity anharmonicity is much weaker than that of the qubit, the predicted Stark shifts show only minor differences compared to the conventional model.

\end{document}